\documentstyle[prd,aps,floats,psfig]{revtex}

\def\lsim{\mathrel{\lower2.5pt\vbox{\lineskip=0pt\baselineskip=0pt
\hbox{$<$}\hbox{$\sim$}}}}
\def\gsim{\mathrel{\lower2.5pt\vbox{\lineskip=0pt\baselineskip=0pt
\hbox{$>$}\hbox{$\sim$}}}}
\def\wh{{\cal W}}
\def\bh{{\cal B}}
\def\wtu{\wh^{\mu \nu}}
\def\wtd{\wh_{\mu \nu}}

\def\Tr{{\rm Tr}}

\newcommand{\NP}[1]{Nucl.\ Phys.\ {#1}}
\newcommand{\ZP}[1]{Z.\ Phys.\ {#1}}

\newcommand{\PL}[1]{Phys.\ Lett.\ {#1}}

\newcommand{\PR}[1]{Phys.\ Rev.\ {#1}}
\newcommand{\PRL}[1]{Phys.\ Rev.\ Lett.\ {#1}}

\begin{document}
\draft
\input epsf
\renewcommand{\topfraction}{0.8}
\twocolumn[\hsize\textwidth\columnwidth\hsize\csname
@twocolumnfalse\endcsname

\preprint{FTUAM 99/31, hep-ph/9912224}
\title{LHC sensitivity to the resonance spectrum of a minimal
strongly interacting electroweak symmetry breaking sector}
\author{A. Dobado$^a$,  M. J. Herrero$^b$,
 J. R. Pel\'aez$^a$ and E. Ruiz Morales$^b$}
\address{$^a$ Departamento de F{\'\i}sica Te{\'o}rica,
  Universidad Complutense de Madrid, 28040-- Madrid,\ \ Spain}
\address{$^b$ Departamento de F{\'\i}sica Te{\'o}rica,
  Universidad Aut{\'o}noma de Madrid,
  Cantoblanco,\ \ 28049-- Madrid,\ \ Spain}
\date{December 3, 1999} 
\maketitle

\begin{abstract}
We present a unified analysis of the two main production processes 
of vector boson pairs at the LHC, $VV$-fusion and $q \bar q$ annihilation,  
in a minimal strongly interacting electroweak symmetry breaking sector.  
Using a unitarized electroweak chiral Lagrangian formalism and modeling
the final $V_L V_L$ strong rescattering effects by a
form factor, we describe $q\bar q$ annihilation processes 
in terms of the two chiral parameters that govern elastic $V_L V_L$ 
scattering. Depending on the values of these two 
chiral parameters, the unitarized amplitudes may present 
resonant enhancements in different angular momentum-isospin channels. 
Scanning 
this two parameter space, we generate the general resonance spectrum 
of a minimal strongly interacting electroweak symmetry breaking sector 
and determine the regions that can be probed at the LHC.
\end{abstract}

\pacs{PACS numbers: 12.60.Fr, 12.39.Fe,\hspace{.3cm} FTUAM 99/31}

\vskip2pc]

\section{Introduction}

One of the main goals of the LHC will be to bring some light on the
symmetry breaking sector (SBS) of electroweak (EW) interactions. In
spite of the impressive agreement of the present data with the Standard
Model (SM) predictions, the origin of EW symmetry breaking remains unknown 
to a large extent. From direct searches of the SM Higgs boson\cite{MHD}
we know that it has to be heavier than 101 GeV (95\% C.L.), and the fit
to EW data \cite{MHFit} gives a 95 \% C.L. upper bound of 230 GeV.
Concerning alternative SBS scenarios, the EW precision measurements disfavor
the most simple technicolor models\cite{PeTa,PDG}, but the data 
are compatible with a general class of strongly interacting SBS \cite{BFS}.  
One of the most characteristic signals of this type of models
is the enhanced production of longitudinal vector boson 
pairs ($V_L V_L$) at high energy colliders \cite{ET}.

The Electroweak Chiral Lagrangian (EChL)\cite{ABL} provides a general way to
describe the low energy effects of different strongly interacting 
SBS models, which are represented by different values of the 
effective chiral couplings. It is inspired in the Chiral Lagrangian
description of low energy pion dynamics in QCD \cite{Wein,GL}.
However, the perturbative predictions made 
with this effective Lagrangian can only describe accurately EW physics
at low energies. The reason is that EW observables 
are given as a truncated series in powers of the external momenta and, 
therefore, they will always violate the unitarity bounds if we go to 
high enough
energy. In particular, at LHC, the EChL amplitudes involving 
longitudinal gauge bosons will violate unitarity for values of the effective 
couplings in the expected range of 10$^{-2}$ to 10$^{-3}$. Furthermore,
these polynomials in the external momenta will not be able to reproduce
the main feature of this type of models, that is, the poles associated
to possible new heavy resonances generated by the SBS dynamics.

The perturbative EChL predictions can be extended to high 
energy using unitarization methods \cite{DHTT}. The unitarized 
amplitudes for $V_L V_L$ production processes can also reproduce a 
resonant behavior depending on the values of the effective couplings.
Since the effective couplings appearing in $VV$ fusion
are different from those in $q \bar q$ annihilation, 
these two processes will not show 
in general the same pattern of resonances for an arbitrary choice of the
effective couplings. 
However, given that the SBS interactions are strong, their dominant 
effects in both processes
are due to the same strong $V_L V_L$ rescattering. Thus, using
the unitarity relations between $q \bar q$ annihilation 
and $V_L V_L$ fusion, we describe in this paper both processes only
in terms of the two chiral parameters that govern elastic 
$V_L V_L$ scattering.

We start giving in Sec.II a brief overview of the Chiral Lagrangian
description of EW interactions.
We summarize the present experimental bounds on the chiral coefficients 
and discuss which are the relevant ones for the present study.
In Sec.III, we provide a unified unitarized description of
the two main $V_L V_L$ production processes at the LHC, 
showing the spectrum of resonances expected in different regions
of the two parameter space. 
In Sec.IV we apply these techniques to study the effects 
of the scalar and vector resonances of the SBS in the production 
of $ZZ$ and $WZ$ pairs at the LHC. First, we briefly describe our 
calculation of the signal and the main background reactions. In order to 
obtain conservative predictions, we will restrict our analysis to the
cleanest detection modes with the final $W$ and $Z$ bosons decaying
into leptons ($e,\mu$). Finally, in Sec. IV, we perform
a systematic study of the significance of the signals for 
both vector and scalar resonances, and determine the region of the
parameter space where these resonances can
be probed at the LHC. We give our conclusions in Sec. V.  

\section{Chiral Lagrangian description of electroweak interactions}

In this work we study minimal strongly interacting symmetry breaking
sectors (MSISBS), in which the global symmetry breaking pattern
$SU(2)_L\times SU(2)_R$ down to the custodial $SU(2)_{\rm c}$ 
symmetry is the smallest one ensuring \cite{Rho} 
that $\rho \approx 1 + {\cal O}(g^2)$ \cite{VeRho}. 
The only light modes of the SBS are the three Goldstone bosons (GB)
associated to this global symmetry breaking.
The next physical states from the SBS are expected to be heavy 
resonances at the TeV scale, generated by the strong interaction dynamics. 
Since no additional Higgs field is included in this approach, the symmetry 
has to be realized nonlinearly, with the three GB, $\omega^a(x)$ with
$a= 1,2,3$, gathered in an SU(2) matrix
\begin{equation}
U(x)=\exp\left(\frac{i\omega^a(x)\tau^a}{v}\right),
\end{equation}
where $\tau^a$ are the Pauli matrices and  $v=246\;\hbox{GeV}$.

The EW interactions at low energies can be well described 
by the Electroweak Chiral Lagrangian \cite{ABL}, an effective 
field theory that couples the three GB to the gauge bosons and 
fermions in an $SU(2)\times U(1)$ invariant way.
This Lagrangian has a set of effective operators of increasing dimension
that represent the low energy effects of the 
underlying symmetry breaking dynamics. 
The C and P invariant bosonic operators up to dimension 4 are
\begin{eqnarray} 
 {\cal L}_{\rm EChL} &=& \frac{v^2}{4} \Tr D_{\mu}U (D^{\mu}U)^\dagger 
                 + a_0\frac{g'^2v^2}{4}[\Tr(TV_{\mu})]^2 
                            \nonumber \\  
                 &+& a_1\frac{i g g'}{2} B_{\mu\nu} \Tr (T \wtu ) 
                 + a_2\frac{ i g'}{2} B_{\mu\nu} \Tr (T[V^{\mu},V^{\nu}])
                            \nonumber \\ 
                 &+& a_3 g \Tr(\wtd [V^{\mu},V^{\nu}])  
                 + a_4 [\Tr(V_{\mu}V_{\nu})]^2 
                               \nonumber \\
                 &+& a_5 [\Tr(V_{\mu}V^{\mu})]^2   
                 + a_6 \Tr(V_{\mu}V_{\nu})\Tr(TV^{\mu})\Tr(TV^{\nu}) 
                              \nonumber \\ 
                 &+& a_7 \Tr(V_{\mu}V^{\mu}) [\Tr(TV^{\nu})]^2 
                 + a_8\frac{g^2 }{4} [\Tr(T \wtd )]^2 
                                 \nonumber\\ 
                 &+& a_9 \frac{g}{2}\Tr(T \wtd )\Tr(T[V^{\mu},V^{\nu}]) 
                                  \nonumber\\ 
                 &+& a_{10}[\Tr(TV_{\mu})\Tr(TV_{\nu})]^2 
                                 \nonumber \\                  
                 &+& \hbox{e.o.m. terms} + \hbox{standard YM terms} 
\label{lag}               
 \end{eqnarray} 
where the ``e.o.m'' terms refer to other operators that can be removed using the
equations of motion, the ``standard YM terms'' stand for 
the gauge fixing and Faddeev-Popov terms, and we have defined the following 
combinations of fields
\begin{equation}
 T \equiv U \tau^3 U^\dagger; \;\;\;\; V_\mu \equiv (D_{\mu}U)U^{\dagger} 
\end{equation}
and the covariant derivative and field strength tensors are given by
\begin{eqnarray}
 D_\mu U &\equiv &\partial_\mu
U - g \wh_\mu U + g' U \bh_\mu , \nonumber\\ 
\wh_\mu  &\equiv &\frac{ -i}{2}\; \vec{W}_\mu \cdot \vec{\tau} ,  \;\;\;\;
\bh_\mu \equiv  \frac{ -i}{2} \; B_\mu \;
\tau^3, \nonumber\\
\wtd  & \equiv & \partial_\mu \wh_\nu - \partial_\nu \wh_\mu -
g [ \wh_\mu, \wh_\nu ], \nonumber\\ 
B_{\mu\nu} &\equiv&  \partial_\mu B_\nu - \partial_\nu B_\mu .
\end{eqnarray}
The first term in Eq.(\ref{lag}) has the form of a gauged non-linear 
sigma model (NL$\sigma$M) and is universal, giving the mass of the 
$W$ and $Z$ bosons and the ``Low Energy Theorems'' for
longitudinal gauge boson scattering \cite{LET}.
The other operators have model-dependent effective couplings $a_i$
that play a double role.
First, six of them (from $a_0$ to $a_5$) are needed as counterterms to 
cancel the divergences generated in a one-loop calculation with 
the Higgs-less NL$\sigma$M Lagrangian. Note that these divergences
are universal. Therefore, after renormalization, these effective 
couplings will have a logarithmic dependence on the scale that is 
universal \cite{Wein,GL,ABL,HeRu}, and a finite piece 
that depends on the prescription used to renormalize the
NL$\sigma$M divergences.
In this work, we will use the ${\overline{MS}}$ renormalization 
prescription for the effective couplings $a_i$.
Second, the $a_i$ coefficients parameterize the low energy
effects of the underlying symmetry breaking dynamics.
For some particular models of strong symmetry breaking, the values of
these effective couplings can be calculated by integrating out the 
heavy degrees of freedom of the underlying theory, as it has been
done for the SM with a heavy Higgs \cite{HeRu}.  In addition
they have also been calculated  in the large $N_{TC}$ limit of
technicolor models\cite{TCestimates}, as well as for chiral models
within the resonance saturation hypothesis \cite{GL,saturation,Donoghue}.
In all these cases, and for
masses of the heavy resonances in the TeV range, 
the typical size of these effective couplings in the ${\overline{MS}}$
scheme lies in the range $10^{-2}$ to $10^{-3}$. 

The EChL formalism has been applied to constrain 
the effective couplings from EW low energy data. 
For instance, the couplings $a_0, a_1$ and $a_8$ 
contribute to the gauge boson self energies up to order $q^2$ \cite{DHE},
and are related to the T,S and U parameters \cite{HeRu,PeTa,HE} 
as explained below. 
A one-loop EChL calculation of the self-energy combinations entering the 
definition of S, T and U gives
\begin{eqnarray}
S &=& \frac{4 e^2}{ \alpha} \left[ \Pi'_{33}(0) -\Pi'_{3Q}(0) \right]
\nonumber\\
 &=& 16 \pi [ - a_1^{\overline{MS}}(\mu) + 
({\rm NL{\sigma}M-loops})(\mu)],
\nonumber\\
T &=& \frac{ e^2}{ \alpha \  s^2 \ c^2  \ m_Z^2} \left[ \Pi_{11}(0) 
-\Pi_{33}(0) \right] \nonumber\\
 &=& \frac{8 \pi}{c^2} [  a_0^{\overline{MS}}(\mu) + 
({\rm NL{\sigma}M-loops})(\mu)],
\nonumber\\
U &=& 4 \frac{e^2}{\alpha} \left[ \Pi'_{11}(0) -\Pi'_{33}(0) \right]
\nonumber\\
 &=& 16 \pi [ a_8 + ({\rm NL{\sigma}M-loops})].\label{STUself}
\end{eqnarray}
where we have made explicit the $\mu$-dependence of the effective couplings
$a_0$ and $a_1$, and of the contributions from the NL$\sigma$M loops.
Note that, once a renormalization prescription is chosen
(${\overline{MS}}$ in our case), the scale dependence of the 
$NL\sigma$M loops is canceled by the renormalized couplings, 
so that these self-energy combinations are scale and renormalization
prescription independent. 
Eqs.(\ref{STUself}) provide the value of the S,T,U self-energy combinations
in a given model characterized by the values of the 
$a_0, a_1$ and $a_8$ couplings and, in particular, using the values 
calculated in \cite{HeRu}, they give the SM predictions for a heavy 
Higgs boson. Following the original paper by Peskin and Takeuchi\cite{PeTa},
we now define  $\Delta S, \Delta T, \Delta U$ as the differences between the 
vacuum polarization effects in an underlying theory of EW symmetry breaking 
and those in the Standard Model with a reference value of the Higgs mass $m_H$.
We can then use Eqs.(\ref{STUself}) to evaluate both the contributions
from the underlying theory and the SM (provided that
a relatively high Higgs reference mass $m_H \gg M_W$ is chosen), 
and obtain 
\begin{eqnarray}
\Delta S & =& S(a_i) - S_{\rm SM}(m_H) = \nonumber\\
 &=& 16 \pi \left[ - a_1^{\overline{MS}}(\mu) + \frac{1}{12}\frac{5/6-\log m_H^2/\mu^2}
{16 \pi^2} \right], \nonumber\\
\Delta T &=&  T(a_i) - T_{\rm SM}(m_H) = \nonumber \\
 &=& \frac{8 \pi}{c^2} \left[  a_0^{\overline{MS}}(\mu)  - 
\frac{3}{8}\frac{5/6-
\log m_H^2/\mu^2}{16 \pi^2} \right], \nonumber\\
\Delta U &=&  U(a_i) - U_{\rm SM}(m_H) 
 =   16 \pi a_8.  \label{STUDeltas}
\end{eqnarray}
These expressions relate 
the measured $\Delta S, \Delta T,\Delta U$ values obtained
from a fit of the Z-pole observables to the SM with the reference value 
$m_H$ with the chiral effective couplings of the underlying theory. 
The latest fit \cite{PDG} to EW data with $m_H=300$ GeV, gives 
\begin{equation}
\Delta S = -0.26 \pm 14,  
\Delta T = -0.11 \pm 16,  
\Delta U = 0.26 \pm 24,  \label{STUfit}
\end{equation}
which imply the following bounds for the three chiral couplings
\begin{eqnarray}
a_0^{\overline{MS}}(\rm{1 TeV})& = & (4.3 \pm 4.9)\times 10^{-3}, \nonumber\\
a_1^{\overline{MS}}(\rm{1 TeV}) & = & (6.8 \pm 2.8)\times 10^{-3}, \nonumber\\
a_8 & = & (4.9 \pm 4.7)\times 10^{-3}. \label{afit}
\end{eqnarray}
Similar bounds have been obtained in \cite{BFS}, where a higher
$m_H$ reference value has been used in the fit.
These results disfavor the simplest models of strong SBS, like
a heavy SM Higgs boson and rescaled-QCD technicolor models.
Indeed, it has been shown \cite{PeTa} that models with
exact custodial symmetry, a dominance of vector resonances, and whose
underlying SBS dynamics satisfies the
Weinberg sum rules \cite{Weisum}, give a negative
contribution to $a_1$ (that is, a positive contribution 
to S) that is clearly disfavored by the data. 
However, the effective couplings in Eq.(\ref{afit}) are perfectly 
compatible with the general hypothesis of a strong SBS \cite{BFS}, 
because their values are in the expected range 
and no fine tuning is needed in order to fit the data.
The open question is then whether there is a model of underlying SBS dynamics
that can explain these values. In this work, we take a phenomenological 
approach without making any assumption on the underlying theory, 
and investigate what can we expect at future colliders if the EChL couplings
take natural values in the range 10$^{-2}$ to 10$^{-3}$.

At LEP-II and Tevatron, three more effective couplings 
$a_2,a_3$ and $a_9$ come into play, through their contribution 
to the triple gauge boson vertices.
A complete 1-loop EChL calculation \cite{HE} and a fit to the data
could place constraints on these new couplings, but this analysis
has not been done so far. In spite of that, indirect bounds 
\cite{Tril,concha} of the 
order of 10$^{-1}$ for $a_2,a_3$ and $a_9$
and in the range of 10$^{-1}$ to 10$^{-2}$ for 
$a_4, a_5, a_6, a_7$ and $a_{10}$ that contribute to the quartic gauge 
boson vertices, 
can be obtained from the low energy data through their contribution
to anomalous vertices in 1-loop calculations.

To summarize, the EW interactions in a MSISBS
can be well described at low energy by the EChL, 
with a set of effective couplings taking values 
in the range of $10^{-2}$ to $10^{-3}$. 
The signals at low energy are expected to be small deviations
in the EW observables, of a similar size to the EW
radiative corrections.   

Concerning the LHC, there are already studies of its sensitivity
to the $W$ and $Z$ interactions within the 
non-resonant EChL approach \cite{noreson,Belyaev}.
Hence, they are limited to moderate energies, due
to the breaking of unitarity already mentioned in the introduction.
There is a general agreement that, although the
present bounds could be significantly improved, with
these non-resonant studies the LHC would be 
hardly sensitive to values of the chiral
parameters down to the $10^{-3}$ level.
Our aim in this work is to extend these studies 
to include resonances without leaving the EChL formalism.
At the next generation of colliders, we 
will be probing the $W$ and $Z$ interactions at TeV energies,
where the longitudinal components of the weak bosons
behave as their corresponding GB. Since the
GB are modes of the SBS, their self-interactions
are strong and it is reasonable to expect that they will dominate
the standard EW corrections. This allows us to simplify further
the description of the strong SBS effects at high energies.

First, since we are assuming that the SBS interactions preserve 
the custodial $SU(2)_{L+R}$ symmetry, only those operators
that are custodial symmetric (once the gauge interactions are switched off)
can be generated by pure strong interaction effects, and they are expected
to be the relevant ones at high energy.
These are the universal term and the operators corresponding to the $a_i$
couplings with $i=3,4,5$. The couplings of the custodial 
breaking operators should
be generated with at least a partial contribution 
from the $U(1)_Y$ gauge interaction or other sources of custodial breaking, 
that we are assuming to be subleading compared with the strong SBS dynamics.  

It is possible to reduce further the number of operators needed
to describe the dominant effects of the strong SBS interactions 
at high energy colliders. If the strong SBS interactions 
dominate the EW physics at high energy, the key reaction is the scattering
of longitudinal vector bosons, because it can take place through a pure 
strong interaction amplitude. Then, if we know the scattering
amplitudes of longitudinal vector bosons in all the relevant channels, 
this characterizes the main effects of the strong dynamics.
In particular, the main corrections to the EW production
of $V_L V_L$ pairs will be due to their strong rescattering effects \cite{Pe},
and if inelastic channels are neglected, 
we can parameterize all the electroweak $V_L V_L$ production mechanisms 
in terms of only two effective couplings $(a_4, a_5)$ that govern 
the elastic $V_L V_L$ amplitudes.  
We discuss in the next section how to make this parameterization
in the unitarized-EChL formalism. Note that for the rest of the paper
we will drop the $\overline{MS}$ superscripts. 

\section{Unitarization and resonances in the SBS}

\subsection{Elastic $V_LV_L$ scattering}

At high energies, the scattering amplitudes 
of longitudinal gauge bosons can be approximated by the
corresponding GB amplitudes using the Equivalence Theorem (ET) \cite{ET}.
At first sight, it may seem that the ET is incompatible with the
use of the EChL, since the ET is valid only at energies $\sqrt s >> M_W$ 
while the EChL is a low energy effective theory.
Nevertheless, it has been shown \cite{ETnosotros} that
there is still a window of applicability for 
the ET together with EChL, valid at lowest order in the weak
couplings, and for small chiral parameters. 
However, in general, if we want to use
the ET at energies larger than, say, 1 TeV, it is
essential that the theory respects unitarity at high energies.
This is an additional reason to use the unitarization 
methods that we discuss in this section.

The accuracy of the ET approximation
was also studied in  \cite{aet}, by comparing the 
$V_LV_L$ scattering cross-section calculated
at tree level with the ${\cal L}_{\rm EChL}$, eq.(\ref{lag}), 
with and without the ET. The difference
between the EChL cross sections calculated with external $V_L$
and those
calculated with external GB's was found to be $O(1\%)$ as soon
as $\sqrt{s}> 500\,$GeV. 
If, in addition, as in the present work,
the GB cross sections  are considered 
at lowest order on the weak couplings, i.e. $O(g^0)$ 
and $O(g'^0)$ for this subprocess,
the previous difference amounts to $O(10\%)$,
for the resonant channels of this paper, which are the
relevant ones here.

Customarily, GB elastic scattering is described in terms of partial wave
amplitudes of definite angular momentum, $J$,
and weak isospin, $I$, associated to the custodial $SU(2)_{L+R}$ group. 
With the EChL, these partial waves, $t_{IJ}$ are obtained 
as an energy (or external momentum) expansion
\begin{equation} 
t_{IJ}(s)=t^{(2)}_{IJ}(s)+t^{(4)}_{IJ}(s)+O(s^3),
\label{expansion}
\end{equation}
where the superscript refers to the corresponding power of momenta. 
The explicit expressions for these GB amplitudes valid up to $O(p^4)$ 
are given in the appendix \cite{antoniomariajo}.
As long as we are working at lowest order in the weak coupling 
constants and we are assuming custodial symmetry in the SBS, 
these amplitudes only depend on the two parameters $a_4$ and $a_5$.

It is easy to check that the EChL amplitudes given in Eqs.
(\ref{expansion}) and (\ref{pw}) do not satisfy the elastic unitarity 
condition 
\begin{equation}
\hbox{Im}\, t_{IJ}(s) =\mid t_{IJ}(s)  \mid ^2\quad \Rightarrow\quad
\hbox{Im}\,\frac{1}{t_{IJ}(s)}=-1,
\label{tunit}
\end{equation}
which is simply the partial wave version of the 
Optical Theorem. However, they satisfy the following  
perturbative relation
\begin{equation}
\hbox{Im} t_{IJ}^{(4)}(s) = \mid t_{IJ}^{(2)}(s)  \mid ^2
\label{pertunit}
\end{equation}
Whereas this condition is approximately equivalent to the exact one
for the relevant energies at LEP, SLC and Tevatron, 
that is definitely not the case in the TeV energy region. 
In general, and for $(a_4,a_5)$ parameters of a natural size, 
$10^{-2}$ to $10^{-3}$,
the unitarity violations cannot be ignored at energies beyond 1 TeV.

To solve this problem, we are going to unitarize the above amplitudes
by means of the Inverse Amplitude Method (IAM) 
\cite{Truong,IAMpiones,IAMdispersive}. 
This method has given remarkable
results describing meson dynamics 
further beyond the perturbative regime, and reproducing the first 
resonances in each $I,J$ channel up to 1.2 GeV
\cite{IAMdispersive,oop}. 
A simple way to understand the IAM is to realize that, as indicated 
in eq.(\ref{tunit}), the imaginary part 
of the inverse elastic amplitude is known exactly at all energies.
As a consequence, any unitary elastic amplitude has to satisfy 
\begin{equation}
\frac{1}{t_{IJ}(s)}=\hbox{Re}\,\frac{1}{t_{IJ}(s)}-i\quad 
\Rightarrow\quad
t_{IJ}(s)=\frac{1}{\hbox{Re}\,t^{-1}_{IJ}(s) - i}.
\end{equation}
Hence, we only need the EChL to approximate the real part of the 
inverse amplitude. Formally: 
\begin{equation}
\hbox{Re}\,t^{-1}_{IJ}= (t^{(2)}_{IJ})^{-1}[1-\hbox{Re} 
t_{IJ}^{(4)}/t^{(2)}_{IJ}+...\,].  
\end{equation}
Then, using eq.(\ref{pertunit}) we arrive at the final expression 
for the unitary amplitudes 
\begin{equation}
t_{IJ}(s)=\frac{t^{(2)}_{IJ}(s)}{1-t_{IJ}^{(4)}(s)/t^{(2)}_{IJ}(s)} 
\label{IAM}
\end{equation}
which are the $O(p^4)$ IAM  partial waves that respect strict elastic 
unitarity at all energies.
Note that the low-energy chiral prediction (\ref{expansion}) is recovered 
if we re-expand (\ref{IAM}) in powers of $s$, so that we have not spoiled 
the good features of the EChL. 

Concerning resonances, although in our derivation of Eq.(\ref{IAM})
we have used Eq.(\ref{tunit}) which only holds for physical values 
of $s$, the very same unitarized amplitudes
can be obtained using dispersion theory \cite{IAMdispersive}, 
thus justifying the extension of eq.(\ref{IAM}) to the complex plane. 
In particular, it can be shown that Eqs.(\ref{IAM}) have 
the proper analytical structure with the right cuts. 
In addition, for certain values of the chiral coefficients, the partial
waves from Eq.(\ref{IAM}) can have poles in the second Riemann sheet,
which can be interpreted as dynamically generated resonances.
Thus within this EChL+IAM formalism one can 
describe resonances without increasing 
the number of parameters and, at the same time, respecting chiral symmetry
and unitarity at all energies.

Note, however, that since the IAM at $O(p^4)$ can only generate
one pair of conjugated poles in the complex-$s$ plane, we can
only reproduce one resonance per channel. Hence, when we identify 
poles with resonances, 
we are implicitly assuming that the values of $a_4$ and $a_5$
describe the GB interactions due to the low energy tail of these 
resonances. The saturation of the chiral
parameters by the lightest resonance multiplets, is usually
known as the resonance saturation hypothesis \cite{GL,saturation},
and the better known strong scenarios are indeed of this type.

Furthermore, non-resonant channels can also 
be well reproduced, since in this case, the IAM poles will appear
at energies so high that the low energy regions look non-resonant.
Although the IAM formula still yields poles, they are beyond the 
applicability limits, where other effects that we are neglecting
here can come into play, and we are not allowed to
interpret them as resonances. 

\subsection{Unitarization of $q\bar{q}\rightarrow V_LV_L$.}

In order to study the LHC sensitivity to the different resonant 
scenarios via $V_L V_L$ production, it is essential to include the  
$q\bar{q}$ annihilation process.
By means of the ET, this process can be estimated from 
$q\bar{q}\rightarrow\omega \omega$. 
Let us recall that the couplings of GB to quarks are proportional to
their mass. Therefore, as far 
as the initial quarks are essentially 
massless the $q\bar{q}\rightarrow z z$ amplitude 
is negligible, and will be ignored.
In addition,  the only relevant contribution to 
$q\bar{q}'\rightarrow w z$ comes from the s-channel, 
where a quark and an anti-quark annihilate producing an
$W$ which gives the $ w z$ GB pair. 
After this initial weak process, we expect that the final state will
re-scatter strongly. In practice, such a $W\rightarrow w z$
interaction can be described with a vector form factor, $F_V(s)$,
by replacing $g$ by $g\, F_V(s)$
(similarly to what happens for the pion form factor).
Due to gauge invariance, $F_V(0)=1$.

The low energy EChL prediction for the form factor is given as a 
series expansion
\begin{equation}
  \label{Fpert}
  F_V(s)=1+ F_V^{(2)}(s)+...
\end{equation}
The explicit EChL expression of $F_V^{(2)}(s)$ is given in the appendix,
but at this moment it is important to note that it depends on the chiral
parameter $a_3$, thus introducing another undetermined constant
in the analysis. 

Since we are only considering strong rescattering effects,
the exact two body unitarity condition for the form factor reads
\begin{equation}
\hbox{Im}\,F_V (s) = F_V(s) t_{11}^*(s),
    \label{Funit}
\end{equation}
Note that, according to our assumption that the strong
SBS interaction preserves custodial symmetry, and due to
the fact that in the final state there are two bosons, there are only
three possible $(I,J)$ elastic scattering channels, 
namely $(0,0), (1,1)$ and $(2,0)$ (as it happens also in pion scattering
in the isospin limit). Consequently, for the vector form factor which
has $J=1$, the final state can only rescatter in the $(1,1)$ channel.

As in the case of the GB elastic amplitudes, the form factor 
in Eq.(\ref{Fpert}) only satisfies unitarity perturbatively, i.e.
\begin{equation}
\hbox{Im}\,F_V^{(2)} (s) = F_V^{(0)}(s) t_{11}^{(2)\,*}(s)= t_{11}^{(2)}(s),
    \label{Fpertunit}
\end{equation}

A way to unitarize the form factor is to realize that the unitarity
condition (\ref{Funit}) tells us that the vector form factor 
$F_V$ should have the same phase and the same poles that the 
$t_{11}$ partial wave.
Therefore, 
\begin{equation}
  \label{samephase}
   \frac{F_V(s)}{t_{11}(s)}= \frac{\hbox{Re}\,F_V(s)}{\hbox{Re}\,t_{11}(s)}.
\end{equation}
Now, we can get an approximation of the modulus of $F_V$ 
using the EChL expressions for $\hbox{Re}\,F_V/\hbox{Re}\,t_{11}$. 
Using the unitarized expression for $t_{11}$ from Eq.(\ref{IAM}),
we ensure that the poles and phase of $F_V$ are correct. 
In summary, we arrive at
\begin{eqnarray}
  \label{Funitaria}
F_V&\simeq& 
\frac{1+\hbox{Re}\,F_V^{(2)}}{1+\hbox{Re}\,t_{11}^{(4)}/t^{(2)}_{11}}\,
\frac{1}{1-t_{11}^{(4)}/t^{(2)}_{11}}.
\end{eqnarray}
At leading order in the chiral expansion, the first fraction in 
Eq.(\ref{Funitaria}) is one, and the next order correction
depends on the parameter $a_3$ through $\hbox{Re} \, F_V^{(2)}$
and on $a_4, a_5$ through the elastic amplitude $t$.
In addition, we are going to show next
that this fraction can be very well approximated
to one if  the same vector resonance
dominates both $F_V$ and $t_{11}$.
On the one hand, the vector form factor can be unitarized using only
its EChL expansion in eq.\ref{Fpert}, as follows:
\begin{equation}
F_V(s)\simeq\frac{1}{1-F_V^{(2)}(s)}.
\end{equation}
(This formula has been tested successfully in pion physics,
see \cite{Truong} and \cite{hannah}).
With this equation it is possible to generate a pole associated
to a vector resonance while keeping the correct low energy behavior,
much as it happened for elastic scattering and eq.(\ref{IAM}).

On the other hand, if such a vector resonance dominates the final rescattering
of the form factor, it should also be present in the $(I,J)=(1,1)$
scattering amplitude.
That is, both the above equation and eq.(\ref{IAM}) should have 
a resonance at the same mass with the same width. 
Thus, together with eqs.(\ref{pertunit}) 
and (\ref{Fpertunit}), which relate the imaginary parts
of $t_{11}^{(4)}$ and $F_V^{(2)}$, we get the following relation
for the real parts around the pole position, $M_V$:
\begin{equation}
\hbox{Re}\,F_V^{(2)}(M_V)=\hbox{Re}\,t_{11}^{(4)}(M_V)/t^{(2)}_{11}(M_V),
\end{equation}
which means that the first fraction in Eq.(\ref{Funitaria}) 
can be set equal to one
as a very good approximation, not only at low energies
but also at all energies, when the resonance dominates the amplitude.

The use of unitarization methods to derive this result
could suggest some arbitrariness. However, the above 
relation can also be recast in terms of chiral parameters,
using the formulas for $t_{11}^{(4)}$ and $F_V^{(2)}$
given in the appendix.
In the $\overline{MS}$ scheme
it reads
\begin{equation}
a_3(M_V)\simeq a_4(M_V)-2a_5(M_V)-\frac{1}{12}\frac{1}{16\,\pi^2},
\end{equation}
which, apart from the small last factor,
is satisfied in $SU(N)$ models at leading order 
in the large $N$ expansion. 
Note that from the strictest point of view of the
effective lagrangian, these three parameters are independent, although
once
one assumes a particular underlying theory or vector dominance,
one could get a relation among them.
Indeed, for general models of vector dominance, it is possible
to estimate the values of the chiral parameters 
\cite{GL,saturation,Donoghue}
for different resonances in terms of their masses and widths. 
Indeed we have  checked that, for typical vector resonance
masses in the 700 to 3000 TeV range, the 
first fraction in Eq.(\ref{Funitaria}) 
ranges between 1.2 and 1.3 for $\sqrt{s}>500$GeV, although
both the numerator and the denominator are much
larger than one. These are only estimates, although
they suggest that with this approximation we would be underestimating the 
amplitude and therefore our conclusions about
the signal would lie on the conservative side.
 
Thus, at least for scenarios with a vector dominance,
the unitarized vector form factor 
is well approximated by 
\begin{equation}
  \label{Fradikal}
F_V(s)\simeq\frac{1}{1-t^{(4)}_{11}(s)/t^{(2)}_{11}(s)}.
\end{equation}
which is completely determined by the unitarized
$t_{11}(s)$ amplitude and depends only on $a_4$ and $a_5$. 
This approach has also been applied to the pion form factor
and it reproduces the $\rho$ correctly \cite{inpreparation}.
\begin{figure}[thbp]
\begin{center}
\hspace*{-.5cm}
\hbox{\psfig{file=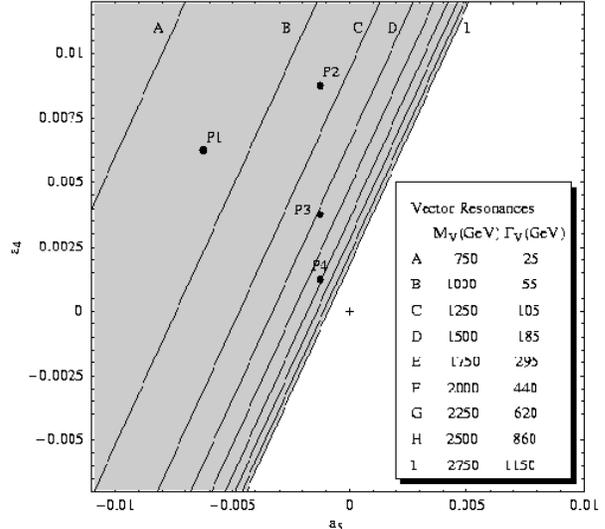,width=8cm}}     
\caption{Vector Resonances in the $(a_4,a_5)$ parameter space.
The chiral couplings are given in the ${\overline{MS}}$-scheme
at the scale of 1 TeV.
The J=I=1 partial wave only depends on $a_4-2a_5$, so that
the straight lines have the same physics in this channel. In the table
we give the resonance parameters for several lines. The points P1 to P5 
will be used as reference models in Sec.IV.}
\label{fig1}
\end{center}
\end{figure}

In models where there is not a vector resonance saturating the
$I=1,J=1$ channel, we do not expect 
a significant enhancement of the vector form factor.

\subsection{Resonances}

The IAM was first applied to the SBS of the EW theory in \cite{DHTT},
to study the signals at the LHC of several specific choices of $a_4$
and $a_5$ that correspond to models with rescaled-QCD  or
Higgs like resonances.
The complete theoretical study of the resonances that are generated
in the $(a_4,a_5)$-plane was performed in \cite{elnene}.
Since we will use this information in the next section, we review 
here the basic results.

\begin{figure}[thbp]
\begin{center}
\hspace*{-.5cm}
\hbox{\psfig{file=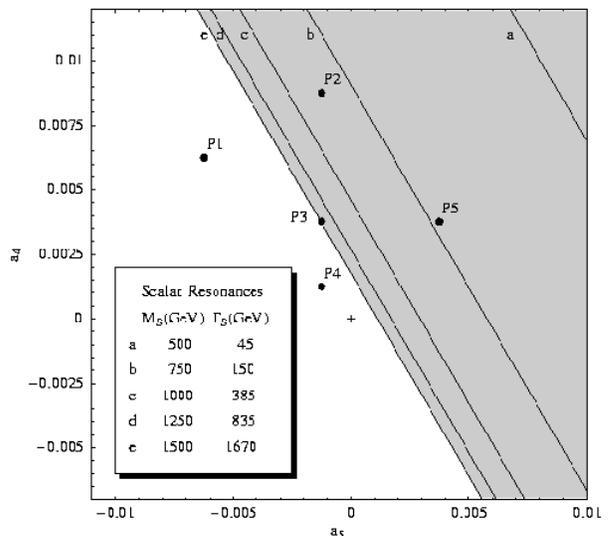,width=8cm}}     
\caption{Scalar neutral resonances in the $(a_4,a_5)$ parameter 
space. The chiral couplings are given in the ${\overline{MS}}$-scheme
 at the scale of 1 TeV.
The J=I=0 partial wave only depends on $7a_4+11a_5$, so that
the straight lines have the same physics in this channel. In the table
we give the resonance parameters for several lines. The points P1 to P5 
will be used as reference models in Sec.IV.}
\label{fig2}
\end{center}
\end{figure}

Scanning the $(a_4,a_5)$ parameter space in the range 
between $10^{-2}$ and $10^{-3}$, we can reproduce the scattering 
amplitudes for $V_L V_L$ production in the MSISBS.
Furthermore, the position of the poles in these amplitudes will give us 
the masses and widths of the resonances 
(see the appendix for the explicit expressions).
We show in Fig.1 a map of the vector resonances
(J=I=1 channel) in the $(a_4,a_5)$ parameter space. 
Within our approximations,
this partial wave only depends on 
the combination $a_4-2a_5$, so that the straight lines with
constant $a_4-2a_5$ have vector 
resonances with roughly the same mass and width. We give several examples
in the table within the figure. In addition we locate five
points that we will use later as illustrative examples.
The area in blank stands for the case when no resonances or saturation 
of unitarity is reached below $4\pi\,v\simeq3\,\hbox{TeV}$, which,
on general grounds, we expect to be the applicability region
of our approach. 
Similarly, we show in Fig.2 the map of neutral scalar resonances 
that appear in the
J=I=0 channel, which only depends on $7a_4+11a_5$
\footnote{J.R.P. thanks J.A.Oller for pointing out a 
mistake in the combination given in \cite{elnene}.
The figures obtained in that paper are nevertheless correct.}. 
Incidentally, the fact that the IAM amplitudes only depend on one 
combination of chiral parameters implies that their mass and width are 
related by the well known KSFR relation \cite{KSFR}.

We do not give the I=2, J=0 channel since we do not expect here 
any resonance in a MSISBS. Intuitively this can be understood from the 
fact that, at low energies, the I=2, J=0 channel is
repulsive and therefore we do not expect doubly charged heavy resonances.
Furthermore, since we cannot make the slope of a phase shift too negative
due to causality (we cannot make an interaction so repulsive that 
the scattered waves leave the interaction point before they arrive),
certain combinations of chiral parameters are excluded theoretically 
\cite{elnene}. Taking all this into account, in the  I=2, J=0 channel
we either find a non-resonant behavior or an smooth and wide saturation 
of unitarity.

\begin{figure}[thbp]
\begin{center}
\hspace*{-.5cm}
\hbox{\psfig{file=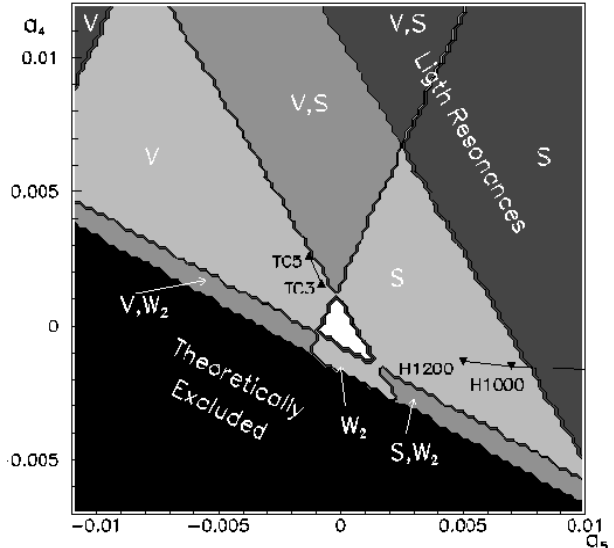,width=8cm}}   
\caption{The general resonance spectrum of a MSISBS in the 
$(a_4,a_5)$ space. The chiral couplings are given in the 
${\overline{MS}}$-scheme at the 
scale of 1 TeV. V stands for vector resonances, S for neutral scalar 
resonances and $W_2$, for wide
structures that  saturate the doubly charged (I=2) channel. 
For illustration, we have also located the most familiar models 
of the SM Higgs and Technicolor, as explained in the text.}
\label{fig3}
\end{center} 
\end{figure}

We have gathered the information on all these channels in Fig.3,
which is a map of the general resonance spectrum of 
a MSISBS\cite{elnene}. Note that depending 
on the parameters, we can find one scalar resonance (S), one vector
resonance (V), two resonances (S,V), a resonance and 
a doubly charged wide saturation effect ($W_2$) or even no 
resonances below 3 TeV (white area).
For illustrative purposes, we have included the 
points that correspond to some simple and familiar scenarios: 
minimal one-doublet technicolor models with 3 and 5 technicolors
(TC3 and TC5), and the heavy Higgs 
SM case, with a tree level Higgs mass of 1000 and 1200 GeV 
(H1000 and H1200). The black region is excluded
by causality constraints on the I=2, J=0 wave.

Note that the chiral couplings $a_4$ and $a_5$ do have a
scale dependence\cite{GL,ABL,HeRu}
\begin{eqnarray}
a_4(\mu) & = & a_4(\mu') - {1 \over 16 \pi^2} {1\over 12} \log 
{\mu^2 \over \mu^{'2} }, \nonumber\\
a_5(\mu) & = & a_5(\mu') - {1 \over 16 \pi^2} {1\over 24} \log 
{\mu^2 \over \mu^{'2} }. \label{run}
\end{eqnarray}
In Figs.(\ref{fig1},\ref{fig2},\ref{fig3}) they are given at the scale 
of 1 TeV. 
Of course, the physical properties of resonances do not change
if we change the scale, but their location in the $(a_4,a_5)$ plane
will be shifted according to the logarithmic
running of the effective couplings given in Eq.(\ref{run}).

Concerning how reliable these predictions are, we should remember 
that we are neglecting  higher order effects
on the weak couplings, gauge boson masses and other 
inelastic channels that could open before 3 TeV. We can only make
a rough estimate of the accuracy of our predictions based in Chiral 
Perturbation Theory and meson dynamics or using specific models. From 
meson-meson scattering, we know that it is possible 
to reconstruct the lightest resonances from the chiral parameters measured 
at low energy to within 10 to 20\% of their actual values.
We also know that inelastic effects due 
to states of more than two GB are highly suppressed up to the chiral
scale (around 3 TeV in our case).
Concerning specific models, we know that we can mimic 
a heavy Higgs scenario or a technicolor scenario within the same range
of accuracy. It is worth noting that we expect the predictions to get 
worse if the resulting resonances become too light.
For instance, it is possible to see that the IAM results deviate by more 
than 20\% from those of the N/D unitarization of a heavy Higgs 
SM if the mass is less than, roughly, 700 GeV
\cite{Oller}. For higher masses the agreement is much better. 
That is why we have darkened the area where ``Light Resonances'' 
(lighter than 700 GeV) appear. The results
in this area should be interpreted very cautiously. Outside this area
we estimate that the predictions of Fig.3 are reliable within, roughly, a 20\%.

\section{Gauge boson pair production at the LHC}

\subsection{Signal and background processes}

The cleanest way to detect $VV$ pairs at hadron 
colliders is through the isolated, high-$p_T$ leptons produced in their 
leptonic decay modes. For this reason, we will restrict our analysis to 
$ZZ$ and $WZ$ production, assuming that their gold-plated decay modes 
$ZZ \to 4l$ and $WZ \to l\nu \ ll$ (with $l=e,\mu$),
can be identified and reconstructed 
with 100 \% efficiency. 
Realistic simulation studies\cite{ATLAS} have shown that the inclusion 
of silver-plated $W^+ W^¯$, $ZZ$ and $W^\pm Z$ events, in which one 
of the gauge bosons decays to jets, can improve the observability of very
heavy scalar and vector resonances respectively. 
However, the study of these channels would require a detailed study of 
QCD backgrounds and jet reconstruction which is beyond 
the scope of this analysis. Therefore our results, based on 
gold-plated events, are rather conservative. 
In addition, since our theoretical scenario does not predict
any resonances in the $I=2$ channel, we have not studied 
like-sign $W^\pm W^\pm$ pair production. Nevertheless, this final state
could be particularly interesting to test non-resonant models\cite{BCHP,CK}, 
due to its small backgrounds.

At LHC, the main production mechanisms of $ZZ$ and $W^\pm Z$ 
pairs are quark-antiquark annihilation and $VV$ fusion processes.
As we explain below, the contribution from non-fusion diagrams with 
bremsstrahlung of the $V$ bosons is expected to be small after kinematical 
cuts, and have not been included in our calculation.
To evaluate $VV$ fusion processes, we use the Effective-W Approximation (EWA)
\cite{EWA} and take the gauge bosons as real with leading-order (LO) 
energy distribution functions. It has been shown by \cite{NLO} that
the LO distributions overestimate the flux of transverse bosons.
Since our signal comes from processes involving longitudinally
polarized bosons, the uncertainty in the fluxes of 
transverse $V$ bosons will only affect the backgrounds from SM $VV$-fusion 
processes, which are probably overestimated.
For the parton distribution functions,
we have used the CTEQ4 set \cite{CTEQ} in all the calculations, 
evaluated at $Q^2= M_W^2$ in $VV$
fusion processes and at $Q^2 = s$ in $q \bar q$ annihilation
and $gg$ fusion processes, with $\sqrt s$ being the total 
center of mass energy of the parton-parton system.

Since we have not included explicitly the decays of the 
final $W$ and $Z$ bosons to leptons in our programs, 
we have used the gauge boson variables to set
event selection cuts. A first event selection criteria to enhance
the strong $V_L V_L$ production signal over the background is to require
high invariant mass $VV$ pairs with small rapidities.
We have applied the following set of minimal cuts:
\begin{eqnarray} \label{cuts}
& & 500 \  {\rm  GeV} \leq \ M_{V_1 V_2} \  \leq 10 {\rm TeV} \nonumber\\[2mm]
& & |y_{\rm lab}(V_1)|, \ |y_{\rm lab}(V_2)|  \ \leq \  2.5 \\[2mm]
& & p_T(V_1), \ p_T(V_2) \  \geq   \ 200  \ {\rm GeV} \nonumber 
\end{eqnarray}
Indeed, these cuts are also required by the approximations that we have
made in our analysis. Given that $V_L V_L \to V_L V_L$ scattering amplitudes
are calculated using the ET, our predictions can only
be applied to $VV$ boson pairs with high invariant mass.
In addition, bremsstrahlung $V$ bosons in non-fusion diagrams
are predominantly produced at small angles, and it is a good 
approximation to neglect their contribution if one restricts the 
analysis to $V$ bosons with high $p_T$ in the central rapidity region. 
Finally, the $p_T$ cut selects $V$ bosons from the signal because they
are produced with high $p_T$ from the two body decay of a 
heavy resonance. However, we should keep in mind that our 
$p_T$ distributions have several sources of uncertainty. 
In $VV$ fusion processes, we have used the EWA assuming collinear $V$
radiation, thus we have neglected the $p_T$ of the incident $V$ bosons. 
In $q \bar q$ annihilation processes,
we have not included the NLO QCD corrections \cite{NLOQCD}, which
are known to increase significantly the distributions at high 
$p_T$ values. In the next section, 
this minimal set of cuts will be complemented with
a more restrictive cut in the invariant mass around the resonances, 
in order to improve the statistical significance of the signal.

The strong-interaction signal in $ZZ$ production is expected in
the fusion channels:
$$
W^+_L W^-_L \to Z_L Z_L,  \ \ \ 
Z_L Z_L \to Z_L Z_L.
$$
The amplitudes for these processes have
been calculated following the approach explained in Sect. III.
We have included  and estimated the following backgrounds:
$$
\begin{array}{ll}
q \bar q \to Z Z, &   61\%  \\[2mm] 
W^+ W^- \to Z Z, &  18\% \\[2mm]
g g \to  Z Z,  &  21\%
\end{array}
$$
where the percentage is their relative contribution to the total background
with the minimal set of cuts. The $ZZ\to ZZ$ background has not
been included since its contribution is known to be negligible
compared with $W^+ W^- \to ZZ$
The continuum from $q \bar q$ annihilation has tree level SM
formulas. As we have said before,
the next to leading order QCD corrections to this process
can significantly enhance the tree level cross sections. 
Therefore, our estimates 
of the $q \bar q$ annihilation background for $ZZ$ production
are probably too optimistic.
The second  background is calculated in the SM at tree level,
with at least one transverse weak boson, excluding the Higgs contribution. 
Finally, the one-loop amplitudes
for $g g \to  Z Z$ have been taken from ref.\cite{GG}. 

For $W^\pm Z$ final states, two processes contribute to the signal:
$$
W^\pm_L Z_L \to W^\pm_L Z_L, \ \ \ 
q \bar q' \to  W^\pm_L Z_L. 
$$
and the backgrounds included in our analysis are
$$
\begin{array}{ll}
W^\pm  Z \to W^\pm Z, & 18\%.  \\[2mm]
\gamma  Z \to W^\pm Z, & 15\%. \\[2mm]
q \bar q' \to  W^\pm Z. &  67\%.
\end{array}
$$
All these backgrounds have SM tree level calculations.
The amplitudes for $W^\pm  Z \to W^\pm Z$ have
at least one transverse weak boson and exclude the Higgs contribution.
In the $q \bar q' \to  W^\pm Z$ background, we do not include
the amplitude with two longitudinal weak bosons, 
which is considered as part of the signal.
The QCD corrections to $q \bar q'$
annihilation processes would give an enhancement
in both the signal and the background, so we expect that they 
will not modify considerably our estimates of the statistical significance 
of vector resonance searches.
We have not studied the contribution to the background from
$t \bar t$ production, but it has been shown that it can be efficiently 
suppressed, after imposing kinematic constraints and isolation cuts to
high $p_T$ leptons\cite{ATLAS}.

\subsection{Numerical Results}

In order to see the LHC sensitivity to the resonance spectrum described
in Sec.III, we have first chosen five 
representative
points in the $(a_4,a_5)$ parameter space 
(see Figs.(\ref{fig1},\ref{fig2})).
Points 1, 3 and 4 represent models containing 
a $J=I=1$ resonance with masses in the range 900-2000 GeV.
Point 5 represents a model with a scalar resonance with  mass
730 GeV and a width of 140 GeV. Finally,
 point 2 corresponds to
a situation with both a scalar and a vector resonance.

\begin{figure}[htbp]
\begin{center}
\hspace*{-.5cm}
\hbox{\psfig{file=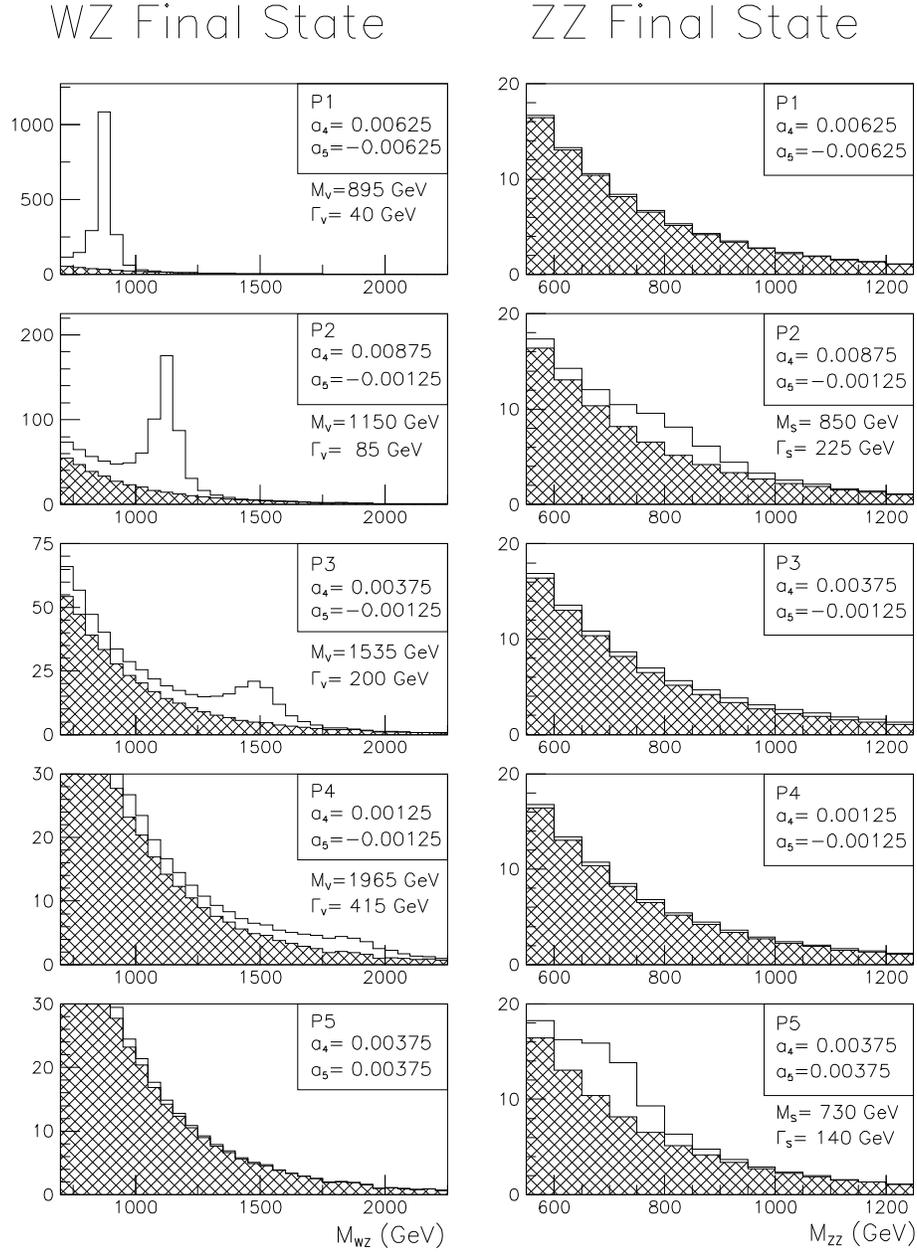,width=12cm}}
\vspace*{.5cm}
\caption{Distribution of gold-plated $WZ$ and $ZZ$ events for 50 GeV 
$M_{VV}$ invariant mass bins, with the minimal set of cuts in 
Eq.(\ref{cuts}). The shaded histogram corresponds to the total background.
On top of it we have plotted the signal as a white histogram. 
We plot (from top to bottom) the predictions for the points
P1 to P5 (see Figs. 1 and 2), that represent models with
one narrow vector resonance, a vector
and a scalar resonance, an intermediate vector resonance, a very wide
vector resonance and, finally, a ``narrow'' scalar resonance.}
\label{fig4}
\end{center}
\end{figure}

The $M_{VV}$ distributions for these five models are shown
in Fig.\ref{fig4}, where we have plotted the signal on top
of the background for gold-plated $ZZ$ and $WZ$ events, assuming
an integrated luminosity of 100 fb$^{-1}$.
The vector resonances in points 1 to 4 can be seen as peaks in
the invariant mass distribution for final $WZ$ states.
The scalar resonances in points 2 and 5 give small enhancements
in the number of $ZZ$ pairs.
We can see that, as $a_4$ and $a_5$
approach the origin, the resonances become
heavier and broader, and therefore the signals in the $M_{VV}$ distributions
are more difficult to detect. 
From these plots, it is also evident that 
that it will be much harder to detect 
scalar than vector resonances. The reasons are: First, that 
scalars are not significantly produced in $q \bar q$ annihilation.
Second, the smaller rate of $ZZ$ production from $VV$-fusion. 
Third, the fact that the branching ratio to leptons 
is smaller for $ZZ$ (BR=0.0044) than for $WZ$ 
final states (BR=0.015), and, finally, that the scalar 
resonances are approximately six times
wider than vector resonances for the same mass.

The relative contribution of the different signal and background processes
for $WZ$ and $ZZ$ production at these representative points
is given in Tables \ref{tab1} and \ref{tab2}.
In order to enhance the signal to background ratio,
we have optimized the cut in $M_{VV}$, keeping events
in the region of approximately one resonance width around
the resonance mass. The $M_{VV}$ cuts taken in each case are given
in the second column of these tables.

From the results for $WZ$ production, it is clear that
the LHC will have an extremely good sensitivity to
light vector resonances, due to their production through 
$q \bar q'$-annihilation which 
dominates by far the $VV$-fusion process.
As the mass of the vector resonance increases, the $q\bar q$ contribution  
is damped faster than the $VV$ fusion, and both 
signals become comparable for vector masses around 2 TeV.
It is also important to note that, in $ZZ$ production, the strong interaction
signal appears only in $VV$ fusion diagrams, and therefore to tag forward jets
is always convenient in this final state in order to reject non-fusion 
processes. This is not the case, however, for vector resonance searches
in $WZ$ pairs because then the most important contribution 
comes from $q \bar q$ annihilation processes.
In these tables, we have also estimated the statistical significance of the 
signal defined as $ S/\sqrt B$, assuming integrated luminosities
of 100 and 400 fb$^{-1}$. In $ZZ$ final states, we also give the 
significance of the signal assuming perfect forward jet-tagging.

\section{Conclusions}

We have presented a unified description of 
longitudinal gauge boson pair production  by fusion and 
$q \bar q$ annihilation just in terms of the $a_4$ and $a_5$ 
parameters of the Electroweak Chiral Lagrangian (EChL). Our amplitudes 
respect unitarity and generate dynamically resonances depending
on the values of these parameters. 
Within this approach, we have studied the sensitivity of the
LHC to the general resonance spectrum of the minimal strongly 
interacting symmetry breaking sector. 

From a purely phenomenological EChL approach,
and without making any further assumption on
the underlying symmetry breaking sector dynamics,
 the present bounds on the electroweak
parameters have room for scenarios where heavy
scalar or vector resonances can appear
in longitudinal gauge boson pair production processes.

\begin{figure}[thbp]
\begin{center}
\hspace*{-.5cm}
\hbox{\psfig{file=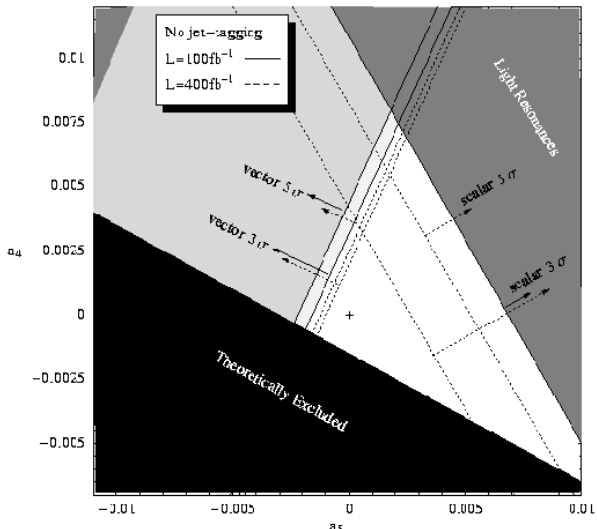,width=8cm}}     
\caption{Sensitivity of the LHC to the resonance spectrum of the strong SBS,
with $WZ$ and $ZZ$ gold plated events.
In the $(a_4,a_5)$ parameter space we represent the $3\sigma$ and 
$5\sigma$ reach with an integrated luminosity of 100 fb$^{-1}$ (solid 
lines limiting the shaded areas)
and 400 fb$^{-1}$ (dashed lines), both for scalar and vector resonances.
The chiral couplings are given in the ${\overline{MS}}$-scheme
at the scale of 1 TeV.}
\label{fig5}
\end{center}
\end{figure}

We show in Figure \ref{fig5} the regions of the 
$(a_4,a_5)$ parameter space that could be tested at the LHC,
giving 3 and 5 sigma contours and assuming integrated luminosities of
100 and 400 fb$^{-1}$. 

We can see that there is a central region in the
$(a_4, a_5)$ parameter space that does not give significant
signals in 
\onecolumn
\twocolumn
\noindent
gold-plated $ZZ$ and $WZ$ events. This region corresponds
to models in which, either the resonances are too heavy to give
a significant enhancement at LHC energies, or there 
are no resonances in the SBS and the scattering
amplitudes are unitarized smoothly.
It is a very important issue whether this type of non-resonant 
$VV$ scattering signals could be probed at the LHC.
Some authors \cite{CK} have argued that doubly-charged $WW$ production
could be relevant to test this non-resonant region.
But non-resonant $VV$ production distributions would have slight 
enhancements in the high energy region, and a very accurate
knowledge of the backgrounds and the detector performance
would be necessary in order to establish the existence of
non-resonant signals over the continuum background.

When the sensitivity contours are translated into resonance mass
reach limits, our results are in good agreement with realistic studies 
at LHC \cite{ATLAS}. We find that, with 100 fb$^{-1}$,
scalar resonances could be discovered (5$\sigma$)
in gold-plated $ZZ$ events up to a mass of 800 GeV 
with forward jet-tagging, and vector resonances could be discovered
using gold-plated $WZ$ events up to a mass of 1800 GeV. 


\section*{Acknowledgements.}

We acknowledge partial support from the Spanish Mi\-nisterio 
de Educaci{\'o}n y Ciencia under CICYT projects 
AEN97-1693, and AEN97-1678. ERM also aknowledges support
from the Spanish AME Foundation.

\section*{Appendix}

The EChL predictions 
\cite{GL,antoniomariajo} for the $V_LV_L$ elastic scattering 
$t_{IJ}$ partial waves, in terms of the $\overline{MS}$
renormalized $a_i(\mu)$ couplings, are 
\begin{eqnarray}
 t^{(2)}_{00}&=&\frac{s}{16\,\pi v^2}, \nonumber\\
 t^{(4)}_{00}&=&\frac{s^2}{64\,\pi v^4}
\left[\frac{16(11a_5(\mu)+7a_4(\mu))}{3}\right. 
\nonumber\\
&+&\left.\frac{1}{16\,\pi^2}\left(\frac{101-50 \log(s/\mu^2)}{9}+4\, 
i\,\pi\right)
\right]. \nonumber\\
 t^{(2)}_{11}&=&\frac{s}{96\,\pi v^2}, \nonumber\\
 t^{(4)}_{11}&=&\frac{s^2}{96\,\pi v^4}\left[4(a_4(\mu)-2a_5(\mu))
+\frac{1}{16\,\pi^2}\left(\frac{1}{9}+\frac{i\,\pi}{6}\right)
\right]. \nonumber\\
 t^{(2)}_{20}&=&\frac{-s}{32\,\pi v^2}, \nonumber\\
 t^{(4)}_{20}&=&\frac{s^2}{64\,\pi v^4}
\left[\frac{32(a_5(\mu)+2a_4(\mu))}{3} \right.
\nonumber\\
&+&\left.\frac{1}{16\,\pi^2}\left(\frac{273}{54}-\frac{20\log(s/\mu^2)}{9}
+i\,\pi\right)
\right]. \label{pw}
\end{eqnarray}
Note that the projection in angular momentum has been defined as
\begin{equation}
t_{IJ}=\frac{1}{64\,\pi}\int_{-1}^1\,d(\cos\theta)\,P_J(\cos\theta)\,
T_I(s,t)\;,
\end{equation}
where $T_I$ is the amplitude in the weak isospin basis.

From these amplitudes, and using eq.(\ref{IAM}), we can
obtain the value of the masses and widths of the resonances
when they appear. We only have to determine the position of
the pole in each channel,  and then to identify
its real and imaginary parts with
the mass and half of the width of the resonance.
Thus, for the vector channel, we find:
\begin{eqnarray*}
  M^2_V=\frac{v^2}{4(a_4-2a_5)+\frac{1}{9(4\pi)^2}},\quad \Gamma_V=
\frac{M^3_V}{96\,\pi\,v^2}
\end{eqnarray*}
Of course,  $M_V$ is an observable and cannot depend on the scale.
Indeed, the $a_4-2a_5$ combination is scale independent (see eq.(\ref{run})).
For the scalar channel we get a trascendental equation
\begin{eqnarray*}
  M^2_S=\frac{12\,v^2}{16\left(11a_5(M_S)+7a_4(M_S)\right)
+\frac{101}{3(4\pi)^2}},\quad \Gamma_S=\frac{M^3_S}{16\,\pi\,v^2}
\end{eqnarray*}
Note that the scale $\mu$ is taken at $M_S$.
From the above equations it is easy to see that, for equal
masses, scalar resonances would be six times wider than
vector resonances.

Finally, we give the expression of the $W\rightarrow w z$
vector form factor
up to next to leading order in the EChL:
\begin{eqnarray*}
&&F_V=1+F_V^{(2)}(s)...\\
&&F_V^{(2)}(s)=\frac{s}{(4\,\pi\, v)^2}\left[64\pi^2a_3(\mu)
-\frac{1}{6}\log\frac{s}{\mu^2}+\frac{4}{9}+i\frac{\pi}{6}\right]
\end{eqnarray*}
where the $a_3(\mu)$ is also given in the $\overline{MS}$ renormalization
scheme. The above equation agrees with the result in \cite{Ultima},
where a different renormalization scheme was used.

\newpage 
\vspace*{4cm}
\newpage

\begin{table}[bht]
\onecolumn
\caption{Expected number of signal and background gold-plated $W^\pm Z$ 
events at the LHC with ${\cal L}$ = 100 fb$^{-1}$, for
 four different values
of $(a_4,a_5)$ that give vector resonances in the $900-2000$ GeV 
mass range.
We have applied the cuts in Eq.(\ref{cuts}), with the optimized
cut in the $VV$-invariant mass indicated in each case. 
The statistical significance of the signal is given also for an
integrated luminosity of 400 fb$^{-1}$. 
\label{tab1}}
\vspace{0.4cm}
\begin{center}
\begin{tabular}{|l|c|cccccc|cc|}
P: $M_V$, $\Gamma_V$ (GeV) & Cuts: & 
Signal & Signal & Signal  &  Backg. & Backg. & Backg. &
$S/\sqrt{B}$ & $S/\sqrt{B}$  \\
$\hspace{4mm} (a_4, a_5)\times 10^3$  & $(M_{VV}^{min}, M_{VV}^{max})$ & 
Fusion & $q \bar q$ & Total  &  Fusion & $q \bar q$ & Total  &
 & (400 fb$^{-1}$) \\ \hline
\begin{tabular}{cl}
P1: & 894, 39 \\
& (-6.25,6.25) \end{tabular}
& (700,1000) & 123 & 1630 & 1743 & 74 & 150 & 224 & 116 & 232 \\
\begin{tabular}{cl}
P2: & 1150, 85 \\
& (-1.25,8.75) \end{tabular}
& (900, 1300) & 65 & 369 & 434 & 50 & 84 & 134 & 37 & 75 \\
\begin{tabular}{cl}
P3: & 1535 , 200 \\
& (-1.25,3.75) \end{tabular}
& (1250, 1700) & 24 & 56 & 80 & 21 & 27 & 48 & 11 & 23 \\
\begin{tabular}{cl}
P4: & 1963 , 416 \\
& (-1.25,1.25) \end{tabular}
& (1500, 2350 ) & 10 & 12 & 22 & 14 & 16 & 30 & 4 & 8
\end{tabular}
\end{center}
\twocolumn
\end{table}
\begin{table}[bhh]
\onecolumn
\caption{Expected number of signal and background gold-plated $ZZ$ events 
at the LHC with ${\cal L}$ = 100 fb$^{-1}$, 
for two representative values of $(a_4,a_5)$ with scalar resonances.
We have applied the cuts in Eq.(\ref{cuts}) with the optimized
cut in the $VV$-invariant mass indicated in each case. 
The statistical significance of the signal is also given for the cases
of ideal forward jet-tagging and for an integrated luminosity of 400 fb$^{-1}$.
\label{tab2}}
\vspace{0.4cm}
\begin{center}
\begin{tabular}{|l|c|ccccc|ccc|}
P: $M_S$, $\Gamma_S$ (GeV) & Cuts: & 
Signal &  Backg. & Backg. & Backg. & Backg. &
$S/\sqrt{B}$ & $S/\sqrt{B}$  & $S/\sqrt{B}$  \\
$\hspace{4mm} (a_4, a_5)\times 10^3$  & $(M_{VV}^{min}, M_{VV}^{max})$ & 
Fusion &  Fusion &  $ gg $ & $q \bar q$ & Total  &
 & (jet-tagging) & (400 fb$^{-1}$) \\ \hline
\begin{tabular}{cl}
P2: & 850, 225 \\
& (-1.25,8.75) \end{tabular}
& (600, 1050) & 15 & 10 & 11 & 34 & 55 & 2 & 5 & 4 \\
\begin{tabular}{cl}
P5: & 750 , 140 \\
& (3.25,3.75) \end{tabular}
& (550, 900) & 21 & 10 & 14 & 39 & 63 & 3 & 6 & 5 \\
\end{tabular}
\end{center}\twocolumn
\end{table}


\begin{thebibliography}{99}

\bibitem{MHD} L3 Collaboration, presentation at LEPC, CERN, November 99,
http://l3www.cern.ch/conferences/talks99.html.

\bibitem{MHFit} M. Swartz, presentation at the Lepton-Photom Symposium.

\bibitem{PeTa} M.E. Peskin and T. Takeuchi, \PRL{65} 964 (1990);
\PR{D46} 381 (1992).  

\bibitem{PDG} \textit{Review of Particle Physics}, Particle Data Group,
{\it Eur. Phys. J.} C 3,1 (1998).

\bibitem{BFS} J.A. Bagger, A.F. Falk and M. Swartz, hep-ph/9908327.

\bibitem{ET} J.M. Cornwall, D.N. Levin and G. Tiktopoulos, Phys. Rev.
{D10}, 1145 (1974);
B.W. Lee, C. Quigg and H. Thacker, Phys. Rev. {D16}, 1519 (1977);
M.S. Chanowitz and M.K. Gaillard,  Nucl. Phys. {B261},  379 (1985).

\bibitem{ABL} T. Appelquist and C. Bernard, \PR{D22}, 200  (1980).\\
A. Longhitano, \PR{D22},  1166 (1980),  \NP{B188}, 118 (1981).

\bibitem{Wein} S. Weinberg, Physica 96A, 327 (1979).

\bibitem{GL} J. Gasser and  H. Leutwyler, Ann. Phys. 158, 142 (1984).

\bibitem{DHTT} A.Dobado, M.J.Herrero and T.N.Truong, \PL
{B235}, 129 (1990); A.Dobado, M.J.Herrero and J. Terr\'on, \ZP
{C50}, 205 (1991); \ZP {C50}, 465 (1991).

\bibitem{Rho}
P. Sikivie, L. Susskind, M. Voloshin, V. Zakharov, Nucl. Phys. 
B173, 189 (1980). 

\bibitem{VeRho} M. Veltman, Acta Phys. Polon. B8, 475 (1977); 
Nucl.Phys. B123, 89 (1977). 


\bibitem{LET} M. Chanowitz, M. Golden and H. Georgi,  \PRL{57} 2344 (1986);
\PR{D36} 1490 (1987).  

\bibitem{HeRu}  M.J. Herrero and E. Ruiz Morales, Nucl.Phys. {B418} (1994) 431,
Nucl.Phys. {B437} (1995) 319; D. Espriu and J. Matias, Phys. Lett.
{B341} (1995) 332; S. Dittmaier and C. Grosse-Knetter,
\NP{B459},497 (1996); A. Nyffeler and A. Schenk, 
Phys. Rev. {D53} (1996) 1494; A. Nyffeler, hep-ph/9907294

\bibitem{TCestimates} T. Appelquist and G.-H. Wu, \PR{D48} (1993) 3235.

\bibitem{saturation}
G. Ecker, J. Gasser, H. Leutwyler, A. Pich and E. de Rafael, 
\PL{B223}, 425 (1989). 
G. Ecker, J. Gasser, A. Pich and E. de Rafael, Nucl.Phys. B321, 311 (1989). 

\bibitem{Donoghue} J.F. Donoghue, C. Ramirez and G. Valencia,
\PR{D39}, 1947 (1989).

\bibitem{DHE} A.Dobado, D. Espriu, M.J. Herrero, Phys. Lett. B255, 405 (1991).

\bibitem{HE} D. Espriu, M.J. Herrero, Nucl. Phys. B373, 117 (1992).

\bibitem{Weisum} S. Weinberg, \PRL{18}, 507 (1967).

\bibitem{Tril} S. Alam, S. Dawson, R. Szalapski. \PR{D57}, 1577 (1998);
J.J. van der Bij, Boris Kastening.  Phys.Rev. D60, 095003,1999.

\bibitem{concha} O.J.P. Eboli, M.C. Gonzalez-Garcia, S.F. Novaes,
Phys. Lett. B 339, 119 (1994); P. Hernandez, J. Vegas, Phys. Lett. 
B307, 116 (1993).

\bibitem{noreson} J. Bagger, S. Dawson and G. Valencia, \NP{B399}, 364 (1993);
A. Dobado and M.T. Urdiales, \ZP{C71}, 659 (1996) ;
A. Dobado, M.J. Herrero, J.R.Pel\'aez, E. Ruiz Morales and M.T. Urdiales
\PL{B352}, 400 (1995);
H.-J. He, Y.-P. Kuang, C.P. Yuan, {\it Mod.Phys.Lett} A11, 3061 (1996);
\PR{D55}, 3038,(1997). 

\bibitem{Belyaev} A.S. Belyaev, O.J.P. Eboli, M.C. Gonzalez-Garcia,
 J.K. Mizukoshi, S.F. Novaes, I. Zacharov, \PR{DD59}, 015022 (1999).

\bibitem{Pe} See, for instance, M. Peskin, SLAC-PUB-5798 (1992).

\bibitem{ETnosotros} H. J. He, Y. P. Kuang, X. Y. Li , Phys. Lett. B329,
278 (1994). A. Dobado and J.R. Pel\'aez, Phys.Lett. B329, 469 (1994);
Nucl. Phys. B425, 110 (1994).

\bibitem{aet} A. Dobado, J.R. Pel\'aez and M.T. Urdiales,  Phys.Rev.
{D56}, 7133 (1997).

\bibitem{antoniomariajo} A.Dobado and M.J. Herrero, \PL{B228}, 425 (1989);
 \PL{B233}, 505 (1989); J. Donoghue and C. Ramirez, \PL{B234}, 361 (1990).

\bibitem{Truong} T. N. Truong, \PRL{61}, 2526 (1988)

\bibitem{IAMpiones} T. N. Truong, \PRL{67}, 
2260 (1991); A. Dobado, M.J.Herrero and T.N. Truong, \PL{B235}, 134 (1990).

\bibitem{IAMdispersive}A. Dobado and J.R. Pel\'aez, \PR{D47}, 4883 (1993); 
\PR{D56}, 3057 (1997).

\bibitem{oop} J.A. Oller, E. Oset and J.R. Pel\'aez, Phys. Rev. Lett. 80, 
3452 (1998); Phys. Rev. D59, 074001 (1999); 
Erratum-ibid. D60, 099906 (1999).
F. Guerrero and J. A. Oller, Nucl. Phys. B537, 459 (1999).

\bibitem{hannah} T. Hannah, \PR{D54}, 4648 (1996).

\bibitem{inpreparation} 
A. Dobado, J. R. Pel\'aez and R.Enriquez-Miranda, in preparation

\bibitem{elnene} J.R. Pel\'aez,  Phys.Rev. D55, 4193 (1997). 

\bibitem{KSFR} K. Kawarabayashi and M. Suzuki, Phys.Lett. B16, 225 (1966)
Riazuddin and Fayyazuddin, Phys. Rev. D47, 1071 (1996)

\bibitem{Oller} J.A. Oller, hep-ph/9908493. To appear in Phys. Rev. D.

\bibitem{ATLAS} ATLAS Technical Proposal, CERN/LHCC/94-43; 
CMS Technical Proposal, CERN/LHCC/94-38; 
ATLAS Technical Design Report, Vol.II, CERN/LHCC/99-15. 

\bibitem{BCHP} V. Barger, K. Cheung, T. Han and R.J.N. Phillips,
Phys. Rev. D42, 3052 (1990).

\bibitem{CK} M. Chanowitz and W. Kilgore, Phys. Lett. B322,147 (1994).

\bibitem{EWA} S. Dawson, Nucl. Phys. B249, 24 (1985). 

\bibitem{NLO} J. Lindfors, Z.Phys. {C28} (1985) 427; 
P.W. Johnson, F.I. Olness and W.K. Tung, Phys. Rev.{D36} (1987) 291.

\bibitem{CTEQ}CTEQ Collaboration, MSUHEP-60426, CTEQ-604

\bibitem{NLOQCD}
J. Ohnemus Phys. Rev. D44 (1991) 1403, D44 (1991) 3477;
S. Frixione et al., Nucl.Phys. B383 (1992) 3;
S. Frixione, Nucl.Phys. B410, 280 (1993).

\bibitem{GG}
E.W.N. Glover and J.J. van der Bij, Nucl.Phys. B321, 561 (1989). 

\bibitem{JPR} I. Josa, F. Pauss and T. Rodrigo, in Proc. LHC Workshop,
Vol.II, Aachen, Oct. 1990, CERN 90-10.

\bibitem{Ultima} J. Bagger, S. Dawson and G. Valencia, hep-ph/9204211,
Fermilab-PUB-92/75-T.

\end{thebibliography}
\end{document}